\documentclass[a4paper,11pt]{article}

\usepackage{jheppub} 
\usepackage{graphicx}
\usepackage{amsmath,amssymb,amsthm,bm,bigdelim,multirow}
\usepackage{color}
\usepackage{hyperref}
\usepackage{dsfont}
\usepackage{url}

\begin{document}

\title{Bond-weighting method for the Grassmann tensor renormalization group}

\author[a]{Shinichiro Akiyama}
	\affiliation[a]{Institute for Physics of Intelligence, University of Tokyo, Tokyo, 113-003, Japan}
    	\emailAdd{akiyama@phys.s.u-tokyo.ac.jp}

\abstract{
Recently, the tensor network description with bond weights on its edges has been proposed as a novel improvement for the tensor renormalization group algorithm. The bond weight is controlled by a single hyperparameter, whose optimal value is estimated in the original work via the numerical computation of the two-dimensional critical Ising model. We develop this bond-weighted tensor renormalization group algorithm to make it applicable to the fermionic system, benchmarking with the two-dimensional massless Wilson fermion. We show that the accuracy with the fixed bond dimension is improved also in the fermionic system and provide numerical evidence that the optimal choice of the hyperparameter is not affected by whether the system is bosonic or fermionic. In addition, by monitoring the singular value spectrum, we find that the scale-invariant structure of the renormalized Grassmann tensor is successfully kept by the bond-weighting technique.
}

\date{\today}

\maketitle

\section{Introduction}
\label{sec:intro}

Tensor renormalization group (TRG) is a typical tensor network algorithm, which enables us to compute the partition function or its path-integral representation via contracting tensors. Since the TRG is a variant of the real-space renormalization group, we can easily access the thermodynamic limit or zero-temperature limit. Moreover, the TRG is essentially free from the sign problem in the stochastic numerical methods, because the TRG does not refer to any probabilistic interpretation for the given Boltzmann weight.
After the original work by Levin and Nave in 2007~\cite{Levin:2006jai}, the algorithm of the TRG has been improved and various types of algorithms have been widely applied not only in the condensed matter community but also in the field of particle physics. For the recent progress in the TRG method and its applications, see Refs.~\cite{Banuls:2019rao,Meurice:2020pxc,Okunishi:2021but,Akiyama:2021nhe,Kadoh:2022loj}, for example. 

Recently, a novel improvement of the TRG algorithm has been proposed by Ref.~\cite{PhysRevB.105.L060402}, which is called the bond-weighted TRG (BTRG). The BTRG is a generalization of the original Levin-Nave TRG by introducing bond weights in tensor network representations. There are mainly two significant features of the BTRG. One is that the bond weight, which is controlled by a single hyperparameter, does improve the accuracy of the TRG with the fixed bond dimension. The other is that the BTRG has a nontrivial fixed-point tensor at criticality with an optimal hyperparameter. The second feature tells us that the renormalization-group transformation considered in the BTRG keeps the scale-invariant structure of the local tensor.  In contrast to the conventional improved algorithm such as the tensor network renormalization~\cite{PhysRevLett.115.180405,PhysRevB.95.045117,PhysRevLett.118.110504}, there is no explicit removal of short-range correlation in the BTRG. Therefore, the bond-weighting technique can be regarded as a new way of improvement in the TRG method. In addition, the bond-weighting technique itself is applicable for any tensor network representations in any dimension. \footnote{So far, the BTRG has been applied to the two-dimensional Ising model~\cite{PhysRevB.105.L060402} and CP(1) model with a $\theta$ term~\cite{Nakayama:2021iyp}. }

In this study, we develop the BTRG for fermions and examine the BTRG in evaluating the Grassmann integrals.
 We investigate whether the two features of the BTRG summarized above are taken over to the Grassmann BTRG algorithm. As shown in previous studies~\cite{Shimizu:2014uva,Shimizu:2014fsa,Takeda:2014vwa,Sakai:2017jwp,Yoshimura:2017jpk,Shimizu:2017onf,Kadoh:2018hqq,Akiyama:2021xxr,Akiyama:2021glo,Akiyama:2020soe,Bloch:2022vqz}, one can successfully reproduce the anti-commutativity of the Grassmann numbers in the TRG algorithms and can directly evaluate the Grassmann integrals without introducing any pseudofermion. This is another advantage of the TRG method over the conventional stochastic methods. However, improvement of the accuracy with the fixed bond dimension has not been adequately discussed so far in the context of the Grassmann TRG algorithms, compared with the normal TRG ones which are usually tested in the spin models. Benchmarking with the two-dimensional massless Wilson fermion, we show that the Grassmann BTRG improves the naive Grassmann TRG and the bond-weighting technique allows us to obtain the scale-invariant Grassmann tensor.
 
This paper is organized as follows. In Sec.~\ref{sec:method}, we briefly review the essence of the BTRG before we extend it to the Grassmann tensor network. We present the results of the benchmark test using the two-dimensional free massless Wilson fermion in Sec.~\ref{sec:results}, where we specify the optimal value of the hyperparameter. We show all the advantages confirmed in the original BTRG are carried over to the Grassmann BTRG, including the scale-invariant structure of the Grassmann tensor. Section~\ref{sec:summary} is devoted to summary and outlook.

\section{Grassmann bond-weighted tensor renormalization group}
\label{sec:method}

\subsection{Brief review of the bond-weighted TRG}

\begin{figure}[htbp]
	\centering
	\includegraphics[width=0.9\hsize]{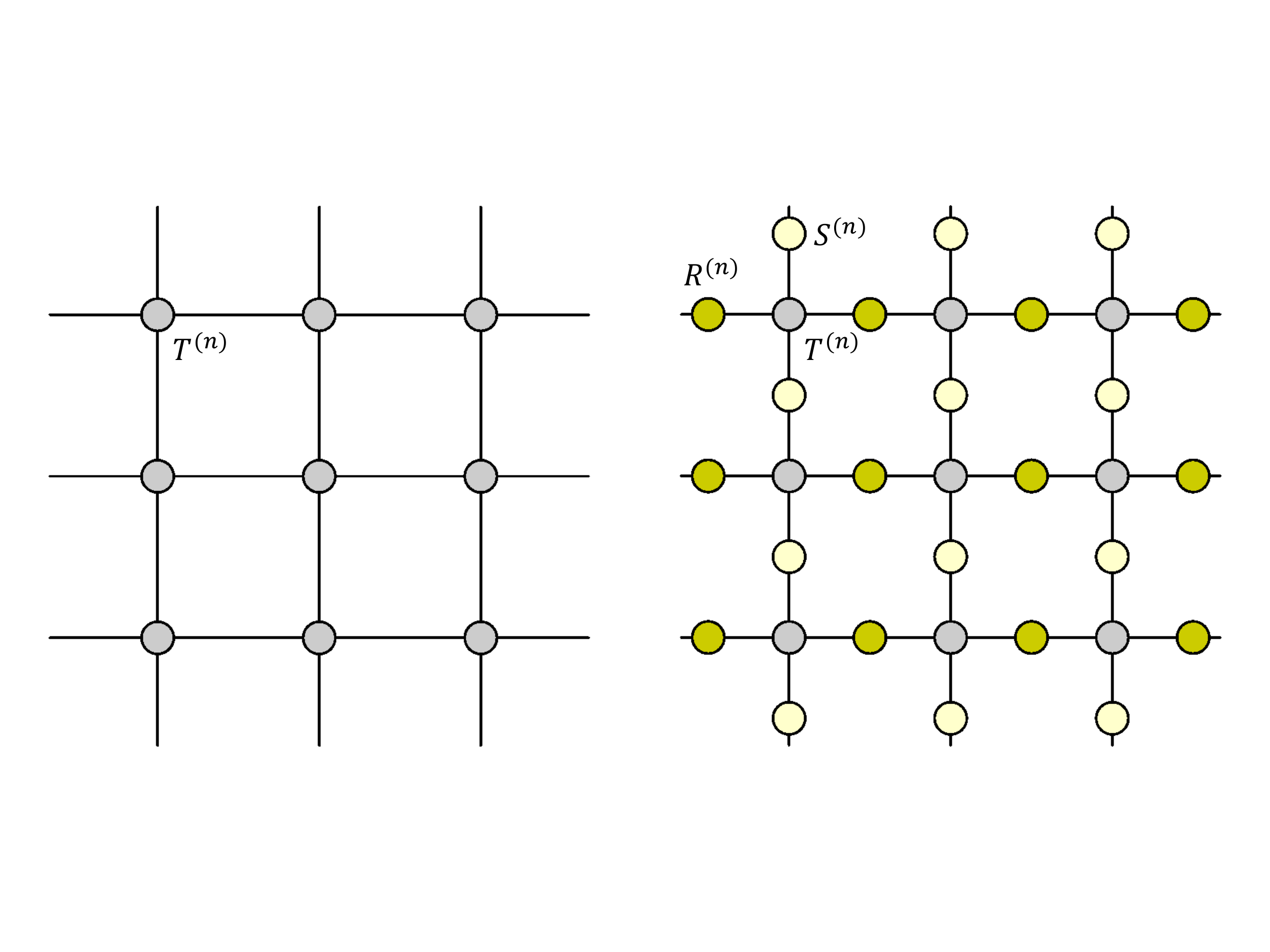}
	\caption{(Left) Tensor network representation considered in the usual TRG algorithms. A local tensor $T^{(n)}$ lives on each lattice site. $n$ denotes the renormalization step. (Right) Tensor network representation considered in the BTRG. Dark and light yellow symbols are the bond weights denoted by $R^{(n)}$ and $S^{(n)}$, respectively, which are associated with the edges of the tensor network. These bond weights are initialized by the identity matrices and updated by the renormalization step of the BTRG.}
  	\label{fig:bond_weight_tn}
\end{figure}

In the BTRG algorithm, we consider a tensor network representation for a partition function or path integral with the bond weights which are located on the edges of the tensor network (Fig.~\ref{fig:bond_weight_tn}). The initial tensor network representation is characterized by the form of
\begin{align*}
	Z={\mathrm{tTr}}\left[\prod_{{\mathrm{lattice~site}}}T^{(0)}\right],
\end{align*}
where $T^{(0)}$ is a four-rank tensor on a two-dimensional square lattice. The BTRG constructs a coarse-graining procedure, which converts $T^{(n)}$ into the coarse-grained tensor $T^{(n+1)}$ where $n$ is the number of coarse-grainings.
The algorithmic essence of the BTRG can be summarized as the singular value decomposition (SVD) with the introduction of a hyperparameter $k\in\mathds{R}$. 
By regarding the four-rank tensor $T^{(n)}$, say $T^{(n)}_{abcd}$, as a matrix $T^{(n)}_{lm}$ where $l$ and $m$ are certain combinations of two subscripts in $T^{(n)}_{abcd}$, we have
\begin{align}
\label{eq:svd}
	T^{(n)}_{lm}
	&=\sum_{i=1}^{{\mathrm{min}}(L,M)}U^{(n)}_{li}\left(\sigma^{(n)}_{i}\right)^{(1-k)/2}\left(\sigma^{(n)}_{i}\right)^{k}\left(\sigma^{(n)}_{i}\right)^{(1-k)/2}V^{(n)*}_{mi}
	\nonumber\\
	&\approx\sum_{i=1}^{D}U^{(n)}_{li}\left(\sigma^{(n)}_{i}\right)^{(1-k)/2}\left(\sigma^{(n)}_{i}\right)^{k}\left(\sigma^{(n)}_{i}\right)^{(1-k)/2}V^{(n)*}_{mi},
\end{align}
where the matrix $T^{(n)}$ is decomposed into the two unitary matrices $U^{(n)}$, $V^{(n)}$ and the singular value $\sigma^{(n)}$. We assign the index $i$ for the singular value.
$L$ and $M$ show the size of subscripts $l$ and $m$, respectively. 
Introducing the bond dimension $D(<{\mathrm{min}}(L,M))$, the row-rank approximation of the matrix $T^{(n)}$ is available. When $k=0$, Eq.~\eqref{eq:svd} corresponds to typical decompositions employed in the original Levin-Nave TRG~\cite{Levin:2006jai}. 
With $k\neq0$, we have the extra factor $(\sigma^{(n)}_{i})^{k}$, which is associated with an edge of the tensor network as we will see below. In Ref.~\cite{PhysRevB.105.L060402}, the authors suggest that the optimal choice of the hyperparameter $k$ be decided by the stationary condition,
\begin{align}
\label{eq:stationary}
	\left[\left(\sigma^{(n)}_{i}\right)^{(1-k)/2}\right]^{4}\left[\left(\sigma^{(n)}_{i}\right)^{k}\right]^{4}=\sigma^{(n)}_{i},
\end{align}
for all $i$ in the BTRG. 
The power of $4$ in Eq.~\eqref{eq:stationary} comes from the fact that renormalized tensor $T^{(n+1)}$ introduced in the BTRG is consist of four unitary matrices, each of them are accompanied with the factor $(\sigma^{(n)}_{i})^{(1-k)/2}$, and four bond weights, each of them are given by $(\sigma^{(n)}_{i})^{k}$, as we will see in Sec.~\ref{subsec:BTRG}.
At each coarse-graining step, the tensor network is rotated by $\pi/4$ and rescaled by a factor of $1/\sqrt{2}$. Therefore, the number of coarse-grainings $n$ provides us with the linear lattice size via $\sqrt{2}^{n}$.
Eq.~\eqref{eq:stationary} should be satisfied in the renormalization of the BTRG if the local tensor $T^{(n)}$ and bond weight $(\sigma^{(n)}_{i})^{k}$ have converged after sufficient times of coarse-graining, sufficiently large $n$ in other words, and if the unitary matrices $U^{(n)}$, $V^{(n)}$ do not affect the singular-value spectrum. 
Based on this assumption, the stationary condition of Eq.~\eqref{eq:stationary} tells us that $k=-0.5$ is optimal. In other words, it is expected that the optimal choice of the hyperparameter just depends on the geometry of the corresponding tensor network. Ref.~\cite{PhysRevB.105.L060402} does confirm that $k=-0.5$ is optimal numerically, benchmarking with the two-dimensional Ising model at criticality.

\subsection{Bond-weighting technique for the Grassmann tensor network}
\label{subsec:BTRG}

We apply the above bond-weighting technique for the Grassmann TRG algorithm. As explained in Ref.~\cite{Akiyama:2020sfo}, the Grassmann TRG evaluates the Grassmann tensor network, which is constructed by the Grassmann tensor whose form is generally given by
\begin{align}
	\mathcal{T}^{(n)}_{\eta_{1}\cdots\eta_{N}}
	=\sum_{i_{1}=0,1}\cdots\sum_{i_{N}=0,1}T^{(n)}_{i_{1}\cdots i_{n}}\eta_{1}^{i_{1}}\cdots\eta_{N}^{i_{N}},
\end{align}
where $T^{(n)}_{i_{1}\cdots i_{N}}$, referred as a coefficient tensor, is a tensor in usual sense and $\eta_{1},\cdots,\eta_{N}$ are the Grassmann numbers. Then, we can decompose $T^{(n)}_{i_{1}\cdots i_{N}}$ as in Eq.~\eqref{eq:svd} by grouping some tensor subscripts to identify it as a matrix. 

\begin{figure}[htbp]
	\centering
	\includegraphics[width=0.9\hsize]{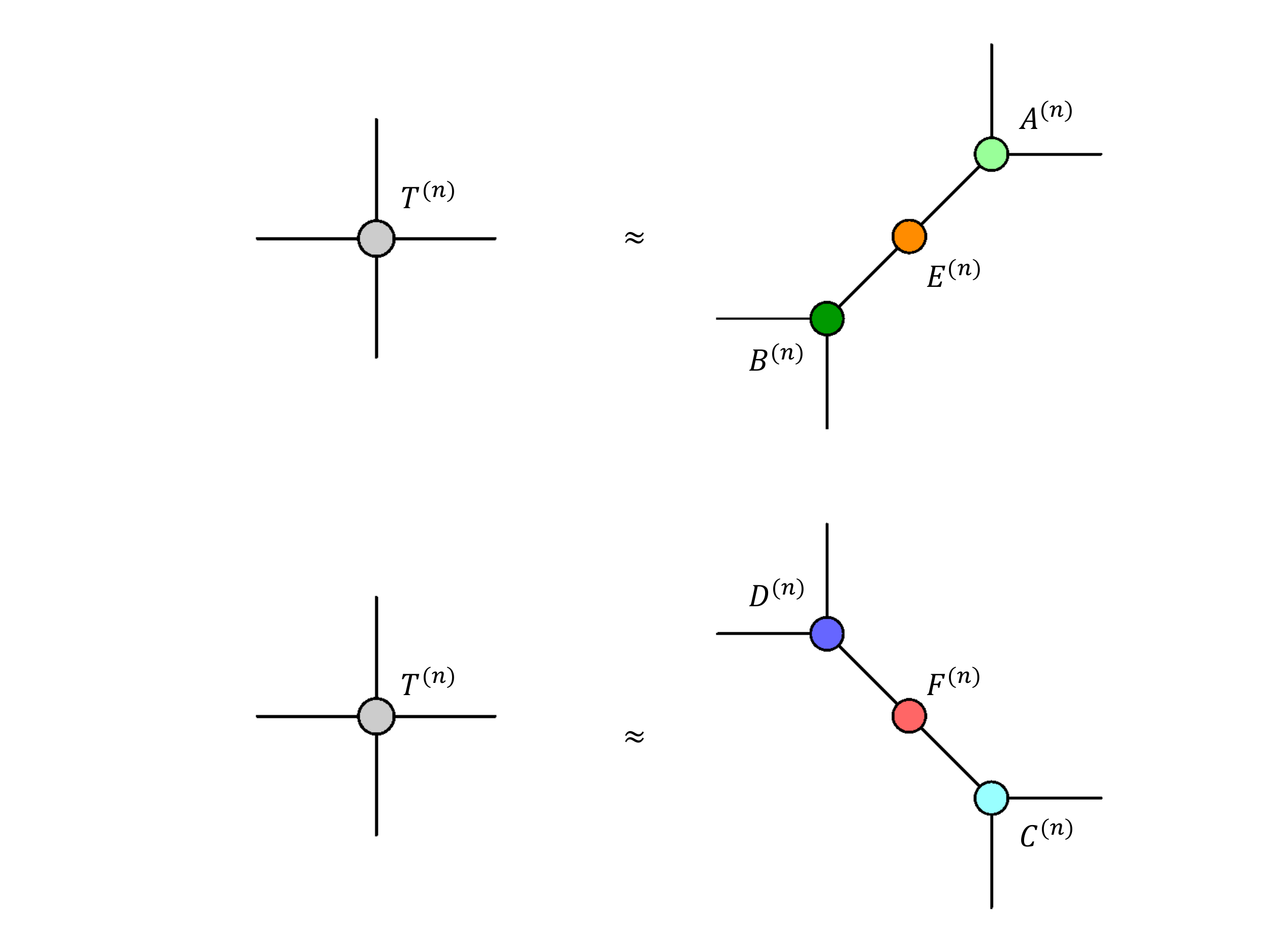}
	\caption{(Top) Diagrammatic representation for Eq.~\eqref{eq:svd_1}. Light and dark green symbols correspond to $A^{(n)}$ and $B^{(n)}$, respectively. The orange symbol shows $E^{(n)}$, which defines the new bond weight $R^{(n+1)}$ via Eq.~\eqref{eq:e_to_r}. (Bottom) Diagrammatic representation for Eq.~\eqref{eq:svd_2}. Light and dark blue symbols correspond to $C^{(n)}$ and $D^{(n)}$, respectively. The red symbol shows $F^{(n)}$, which defines the new bond weight $S^{(n+1)}$ via Eq.~\eqref{eq:f_to_s}. Note that an extra sign factor should be associated with the left-hand side as in Eq.~\eqref{eq:svd_2}.}
  	\label{fig:bond_weight}
\end{figure}

\begin{figure}[htbp]
	\centering
	\includegraphics[width=0.9\hsize]{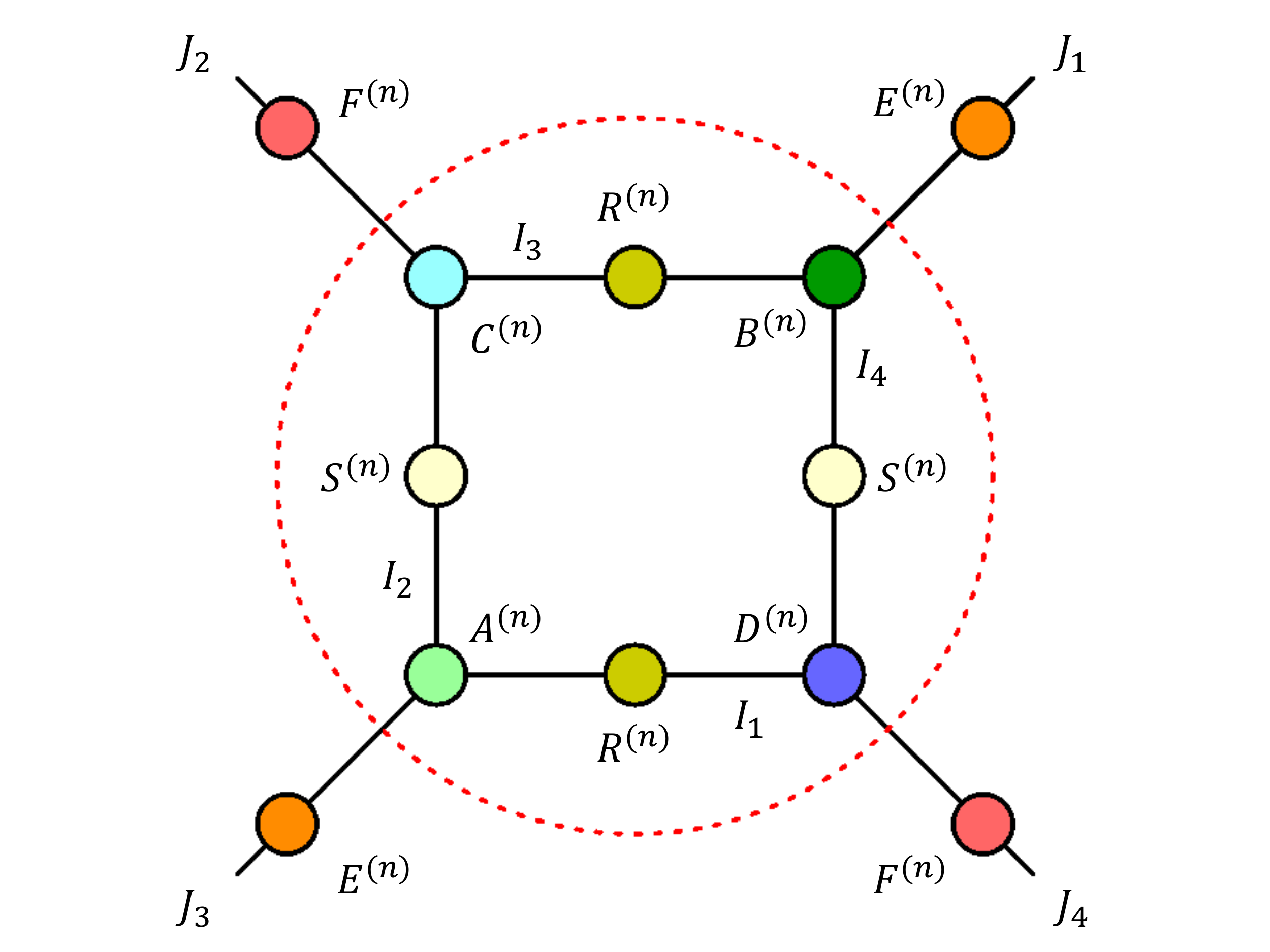}
	\caption{Renormalization step considered in the Grassmann BTRG. The tensor contraction in the red dotted circle shows Eq.~\eqref{eq:renormalization}. Each symbol has already been introduced in Figs.~\ref{fig:bond_weight_tn} and \ref{fig:bond_weight}.}
  	\label{fig:renormalization}
\end{figure}

To obtain the renormalized Grassmann tensor, all we have to do is to carry out the tensor contraction with some sign coming from the Grassmann integrations. To write down this sign factor explicitly, let us introduce a binary function~\cite{Akiyama:2021nhe}. Suppose we have a four-rank Grassmann tensor $\mathcal{T}^{(n)}_{\Psi_{1}\Psi_{2}\Psi_{3}\Psi_{4}}$, which generates the two-dimensional Grassmann tensor network. Generally, the coefficient tensor in $\mathcal{T}^{(n)}_{\Psi_{1}\Psi_{2}\Psi_{3}\Psi_{4}}$ has a structure such as $T^{(n)}_{I_{1}I_{2}I_{3}I_{4}}$ where $I_{\nu}$'s are the sets of binary subscripts, say $I_{\nu}=(i^{(\nu)}_{1},\cdots,i^{(\nu)}_{n})$ with $\nu=1,2,3,4$. $I_{\nu}$ is associated with $\Psi_{\nu}$. A binary function $f_{\nu}$ is defined on each $I_{\nu}$ via
\begin{align}
\label{eq:Gparity}
	f_{\nu}(I_{\nu})
	=
	\begin{cases}
		0 & {\mathrm{if}}~ \sum_{p=1}^{n}i^{(\nu)}_{p}\in2\mathds{Z}\\
		1 & {\mathrm{if}}~ \sum_{p=1}^{n}i^{(\nu)}_{p}\notin2\mathds{Z}
	\end{cases}
	,
\end{align}
which is nothing but the Grassmann parity in the corresponding segment of $\Psi_{\nu}$. \footnote{It is practically useful that we map $I_{\nu}$ into $\tilde{I}_{\nu}\in\mathds{R}$ in advance. Then we define a binary function $f_{\nu}$ on $\tilde{I}_{\nu}$, where $f_{\nu}(\tilde{I}_{\nu})=0$ with $\tilde{I}_{\nu}=1,\cdots,q$, and $f_{\nu}(\tilde{I}_{\nu})=1$ with $\tilde{I}_{\nu}=q+1,\cdots,2^{n}$.} Then we provide the low-rank approximation based on the SVD in the following two ways,
\begin{align}
\label{eq:svd_1}
	T^{(n)}_{I_{1}I_{2}I_{3}I_{4}}
	\approx
	\sum_{J=1}^{D}A^{(n)}_{I_{1}I_{2}J}E^{(n)}_{J}B^{(n)}_{JI_{3}I_{4}},
\end{align}
\begin{align}
\label{eq:svd_2}
	(-1)^{f_{4}(I_{4})(f_{2}(I_{2}+f_{3}(I_{3}))+f_{2}(I_{2})f_{3}(I_{3})}
	T^{(n)}_{I_{1}I_{4}I_{3}I_{2}}
	\approx
	\sum_{K=1}^{D}C^{(n)}_{I_{1}I_{4}K}F^{(n)}_{K}D^{(n)}_{KI_{3}I_{2}},
\end{align}
where $A^{(n)}$, $B^{(n)}$, $C^{(n)}$, $D^{(n)}$ are constructed by unitary matrices and the singular values with the hyperparameter $k$, and $E^{(n)}$, $F^{(n)}$ are defined just by the singular values powered by $k$ (see Eq.~\eqref{eq:svd}) such that
\begin{align}
	A^{(n)}_{I_{1}I_{2}J}=U^{(n)}_{I_{1}I_{2}J}\left(\sigma^{(n)}_{J}\right)^{(1-k)/2},
\end{align}
\begin{align}
	E^{(n)}_{J}=\left(\sigma^{(n)}_{J}\right)^{k},
\end{align}
\begin{align}
	B^{(n)}_{JI_{3}I_{4}}=\left(\sigma^{(n)}_{J}\right)^{(1-k)/2}V^{(n)}_{JI_{3}I_{4}},
\end{align}
\begin{align}
	C^{(n)}_{I_{1}I_{4}K}=U^{(n)}_{I_{1}I_{4}K}\left(\sigma^{(n)}_{K}\right)^{(1-k)/2},
\end{align}
\begin{align}
	F^{(n)}_{K}=\left(\sigma^{(n)}_{K}\right)^{k},
\end{align}
\begin{align}
	D^{(n)}_{KI_{3}I_{2}}=\left(\sigma^{(n)}_{K}\right)^{(1-k)/2}V^{(n)}_{KI_{3}I_{2}}.
\end{align}
Schematic explanation is given in  Fig.~\ref{fig:bond_weight}. $E^{(n)}$ and $F^{(n)}$ directly define new bond weights $R^{(n+1)}$ and $S^{(n+1)}$ via
\begin{align}
\label{eq:e_to_r}
	R^{(n+1)}_{J}=E^{(n)}_{J},
\end{align}
\begin{align}
\label{eq:f_to_s}
	S^{(n+1)}_{K}=F^{(n)}_{K},
\end{align}
expect $n=0$. 
We have assumed that the ordering of subscripts corresponds to the ordering of the Grassmann numbers. This is the reason why we have an extra sign in the left-hand side of Eq.~\eqref{eq:svd_2}, where the corresponding Grassmann tensor is $\mathcal{T}^{(n)}_{\Psi_{1}\Psi_{4}\Psi_{3}\Psi_{2}}$, not $\mathcal{T}^{(n)}_{\Psi_{1}\Psi_{2}\Psi_{3}\Psi_{4}}$. We also note that $J$ and $K$ are associated with new auxiliary Grassmann numbers and we can again introduce binary functions $f_{\mathrm{e}}$ and $f_{\mathrm{o}}$ for $J$ and $K$, respectively. \footnote{As demonstrated in Ref.~\cite{Akiyama:2021nhe}, $f_{\mathrm{e}}$ and $f_{\mathrm{o}}$ are defined via the SVD of a block-diagonal matrix, practically.} The renormalized coefficient tensor $T^{(n+1)}$ is obtained via
\begin{align}
\label{eq:renormalization}
	T^{(n+1)}_{J_{1}J_{2}J_{3}J_{4}}
	&=
	(-1)^{f_{\mathrm{e}}(J_{1})(f_{\mathrm{o}}(J_{2})+f_{\mathrm{e}}(J_{3}))+f_{\mathrm{o}}(J_{2})f_{\mathrm{e}}(J_{3})}
	\sum_{I_{2},I_{4}}
	S^{(n)}_{I_{2}}S^{(n)}_{I_{4}}
	\nonumber\\
	&
	\times
	(-1)^{f_{4}(I_{4})f_{\mathrm{o}}(J_{2})+f_{2}(I_{2})(f_{\mathrm{e}}(J_{3})+f_{\mathrm{o}}(J_{2}))}
	\left\{\sum_{I_{1}}
	(-1)^{f_{1}(I_{1})(f_{2}(I_{2})+f_{\mathrm{e}}(J_{3}))}
	A^{(n)}_{I_{1}I_{2}J_{3}}C^{(n)}_{I_{1}I_{4}J_{2}}R^{(n)}_{I_{1}}
	\right\}
	\nonumber\\
	&
	\times
	(-1)^{f_{4}(I_{4})f_{\mathrm{e}}(J_{1})+f_{2}(I_{2})(f_{\mathrm{e}}(J_{1})+f_{\mathrm{o}}(J_{4}))}
	\left\{\sum_{I_{3}}
	(-1)^{f_{3}(I_{3})(f_{4}(I_{4})+f_{\mathrm{o}}(J_{4}))}
	B^{(n)}_{J_{1}I_{3}I_{4}}D^{(n)}_{J_{4}I_{3}I_{2}}R^{(n)}_{I_{3}}
	\right\}
	.
\end{align}
As remarked in Ref.~\cite{PhysRevB.105.L060402}, $R^{(0)}$ and $S^{(0)}$ should be initialized by identity matrices. 
The above procedure is visually explained in Fig.~\ref{fig:renormalization}.

\section{Numerical results} 
\label{sec:results}

As a benchmark, we evaluate the path integral of the two-dimensional single-flavor massless free Wilson fermion with the Grassmann BTRG. The Grassmann tensor network representation has been already derived in Ref.~\cite{Akiyama:2020sfo}. The Wilson parameter is set to be unity and the anti-periodic boundary condition is assumed in the temporal direction. Since the accuracy of the TRG computation for massless fermions degrades compared with massive ones~\cite{Yoshimura:2017jpk,Akiyama:2020sfo}, we focus on the massless case in the following. 

\subsection{Optimal hyperparameter and bond-dimension dependence}

The action of the single-flavor massless free Wilson fermion on the two-dimensional lattice $\Lambda_{2}=\{(s_{1},s_{2})|s_{i}\in\mathds{Z}~{\rm for}~i=1,2\}$ is given by
\begin{align}
	S&=-\frac{1}{2}\sum_{s\in\Lambda_{2}}\sum_{\nu=1,2}\left[\bar{\psi}(s)(r\mathds{1}-\gamma_{\nu})\psi(s+\hat{\nu})+\bar{\psi}(s+\hat{\nu})(r\mathds{1}+\gamma_{\nu})\psi(s)\right]
	+2r\sum_{s}\bar{\psi}(s)\psi(s),
\end{align}
where $\psi$ and $\bar{\psi}$ are two-component Grassmann fields. $r$ is the Wilson parameter and we set $r=1$ in the following calculations. 
We assume the periodic boundary condition in $1$-direction and the anti-periodic boundary condition in $2$-direction.
As a concrete expression for the two-dimensional $\gamma$-matrix, we employ the Pauli matrix via $\gamma_{1}=\sigma_{x}$ and $\gamma_{2}=\sigma_{z}$. The tensor network representation for the path integral of this action is given by the Grassmann tensor network whose details are demonstrated in Ref.~\cite{Akiyama:2020sfo}. Since the initial bond weights are set to be identity matrices, the Grassmann tensor network derived in Ref.~\cite{Akiyama:2020sfo} is ready to be evaluated by the Grassmann BTRG.

\begin{figure}[htbp]
	\centering
	\includegraphics[width=0.9\hsize]{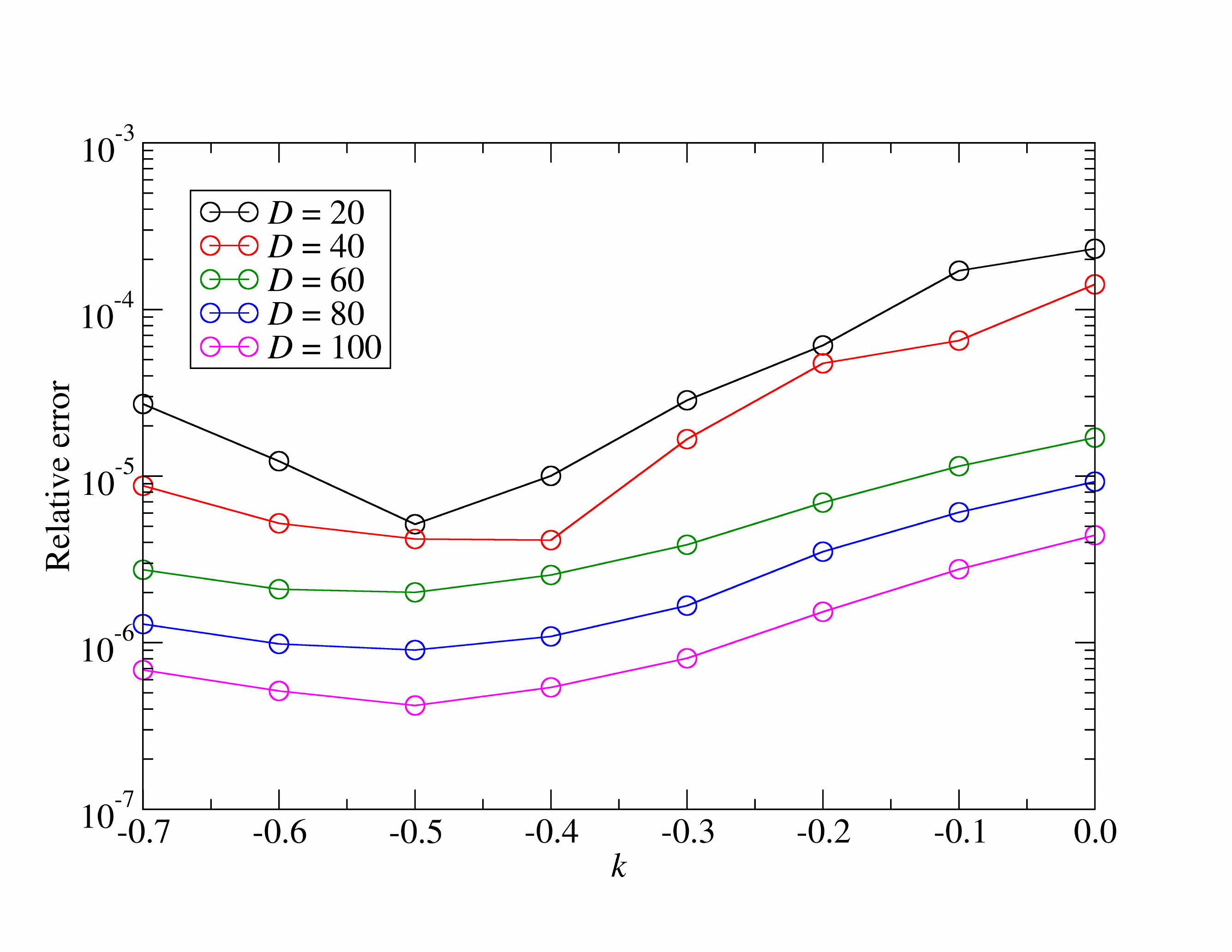}
	\caption{Relative error of the free energy as a function of the hyperparameter $k$. Different curves correspond to the results with the various bond dimension $D$. The relative error is evaluated on the square lattice whose size is $2^{20}$.}
  	\label{fig:k_dep}
\end{figure}

Fig.~\ref{fig:k_dep} shows the relative error of the free energy as a function of the hyperparameter $k$, varying the bond dimension $D$. The relative error is defined on a square lattice whose linear lattice size is $2^{10}$, corresponding to $n=20$. Fig.~\ref{fig:k_dep} suggests that $k=-0.5$ be optimal and this is consistent with the case in the Ising model provided in Ref.~\cite{PhysRevB.105.L060402}. Therefore, the stationary condition in Eq.~\eqref{eq:stationary} is numerically supported not only by the spin model but also by the lattice fermion system. It is worth noting that the accuracy achieved by $D=20,40$ with $k=-0.5$ is higher than that by $D=80$ with no bond-weighting improvement.

\begin{figure}[htbp]
	\centering
	\includegraphics[width=0.9\hsize]{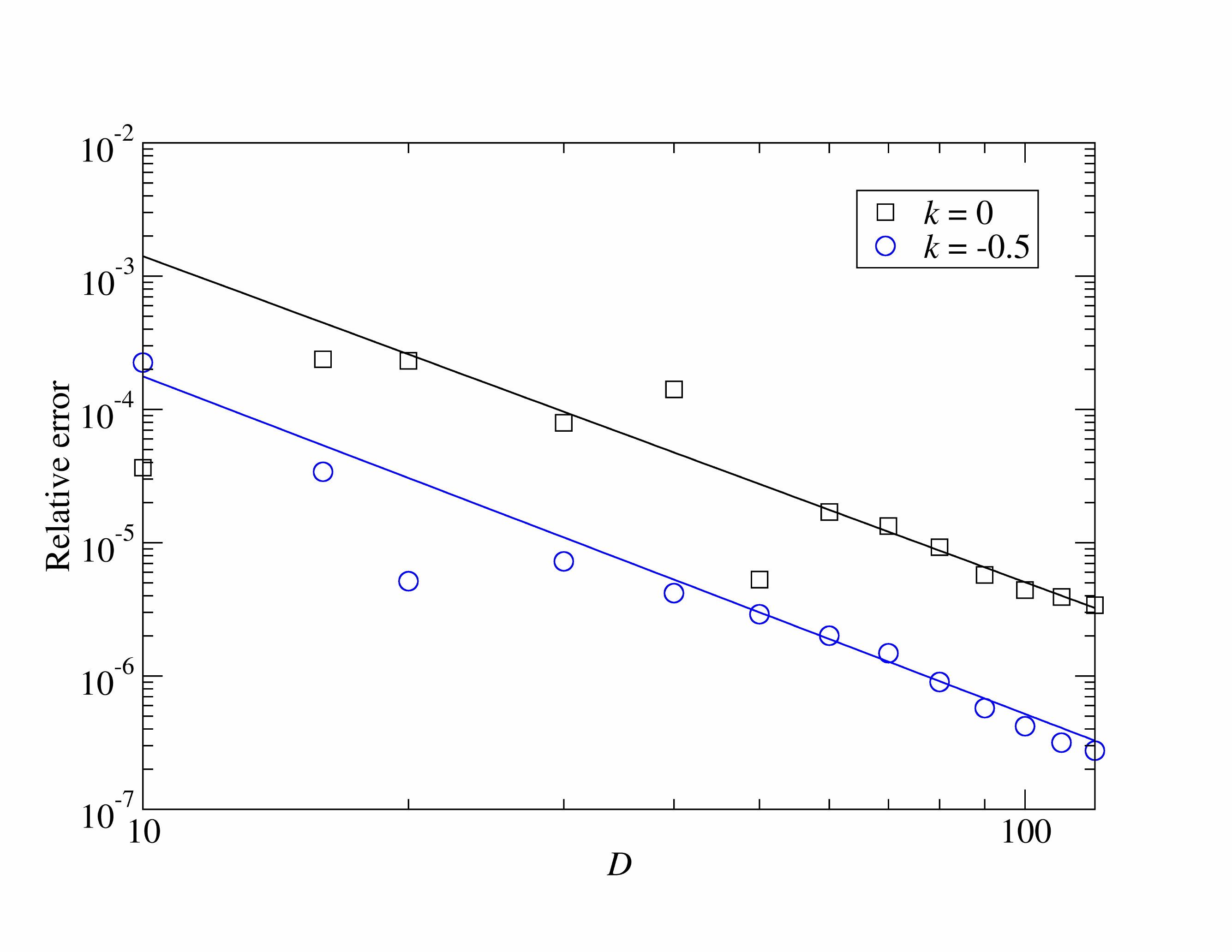}
	\caption{Relative error of the free energy as a function of the bond dimension $D$. The system size is $2^{20}$. The solid lines show the fitting function $aD^{-2\kappa}$, where $a$ and $\kappa$ are fit parameters. Note that $k=0$ corresponds to the Grassmann Levin-Nave TRG and $k=-0.5$ does to the Grassmann BTRG. The fit ranges are set to be $D\in[60,120]$ for $k=0$ and $D\in[50,120]$ for $k=-0.5$, because the Grassmann BTRG yields more stable results for smaller $D$ than the Grassmann Levin-Nave TRG.}
  	\label{fig:d_dep}
\end{figure}

Next, we investigate the relative error as a function of bond dimension, which is shown in Fig.~\ref{fig:d_dep}. The bond dimension is varied up to $D=120$. We compare the relative error obtained by the optimal Grassmann BTRG with that by the Grassmann Levin-Nave TRG, which is characterized by $k=0$. We observe power-law decay in both calculations. According to the so-called finite-entanglement scaling in the matrix product state (MPS)~\cite{PhysRevB.78.024410,PhysRevLett.102.255701}, the effective correlation length $\xi_{D}$ scales with $D^{\kappa}$, where
\begin{align}
\label{eq:kappa}
	\kappa=\frac{6}{c\left(\sqrt{12/c}+1\right)},
\end{align}
and $c$ is the central charge. 
Although the finite-entanglement scaling is established in the MPS, some numerical studies in the Ising model imply that $\xi_{D}\sim D^{\kappa}$ also holds in the TRG algorithms~\cite{PhysRevB.89.075116,PhysRevB.105.L060402}.
Assuming the free energy in the finite-$D$ regime scales with $\xi^{-d}_{D}$ in the $d$-dimensional lattice, one expects that the relative error of the free energy also scales with $\xi^{-d}_{D}$, or $D^{-d\kappa}$.
We extract the exponent $\kappa$ via fitting the data by $aD^{-2\kappa}$, where $a$ and $\kappa$ are the fit parameters, and the result is summarized in Table~\ref{tab:fit}. We find that the Grassmann BTRG results in $\kappa=1.26(7)$ with $a=0.06(4)$ and the Grassmann Levin-Nave TRG does in $\kappa=1.22(8)$ with $a=0.4(3)$. 
The former is closer to $\kappa=1.344\cdots$, which is obtained from Eq.~\eqref{eq:kappa} by setting $c=1$. 
It seems reasonable because the massless Dirac fermion is characterized by $c=1$. 
Since Ref.~\cite{PhysRevB.105.L060402} finds that the BTRG reproduces $\kappa$ in Eq.~\eqref{eq:kappa} in the critical Ising model, which is characterized by $c=1/2$, this is another supporting evidence that the Grassmann BTRG shares the similar converging behavior in terms of $D$ with the original BTRG. 
It should be noticed that the resulting coefficient $a$ is reduced by an order of magnitude when going from the Grassmann Levin-Nave TRG to the Grassmann BTRG, so the relative error is reduced by an order of magnitude.
Fig.~\ref{fig:d_dep} also tells us that the Grassmann BTRG yields more stable results for smaller $D$ compared with the Grassmann Levin-Nave TRG. This is the reason why we have employed different fit ranges to extract $\kappa$ and $a$ for different algorithms as shown in Table~\ref{tab:fit}. 

\begin{table}[htb]
	\caption{Fit result of the exponent $\kappa$. }
	\label{tab:fit}
	\begin{center}
	  	\begin{tabular}{|c|cc|}\hline
			               &  $\kappa$ & fit range\\ \hline  
			Grassmann Levin-Nave TRG ($k=0.0$)  &  1.22(8) & $[60,120]$\\ \hline
			Grassmann BTRG ($k=-0.5$)                  &  1.26(7)  & $[50,120]$\\ \hline
		\end{tabular}
	\end{center}
\end{table}

\subsection{Scale-invariant structure in the Grassmann tensor}
 
\begin{figure}[htbp]
	\centering
	\includegraphics[width=0.9\hsize]{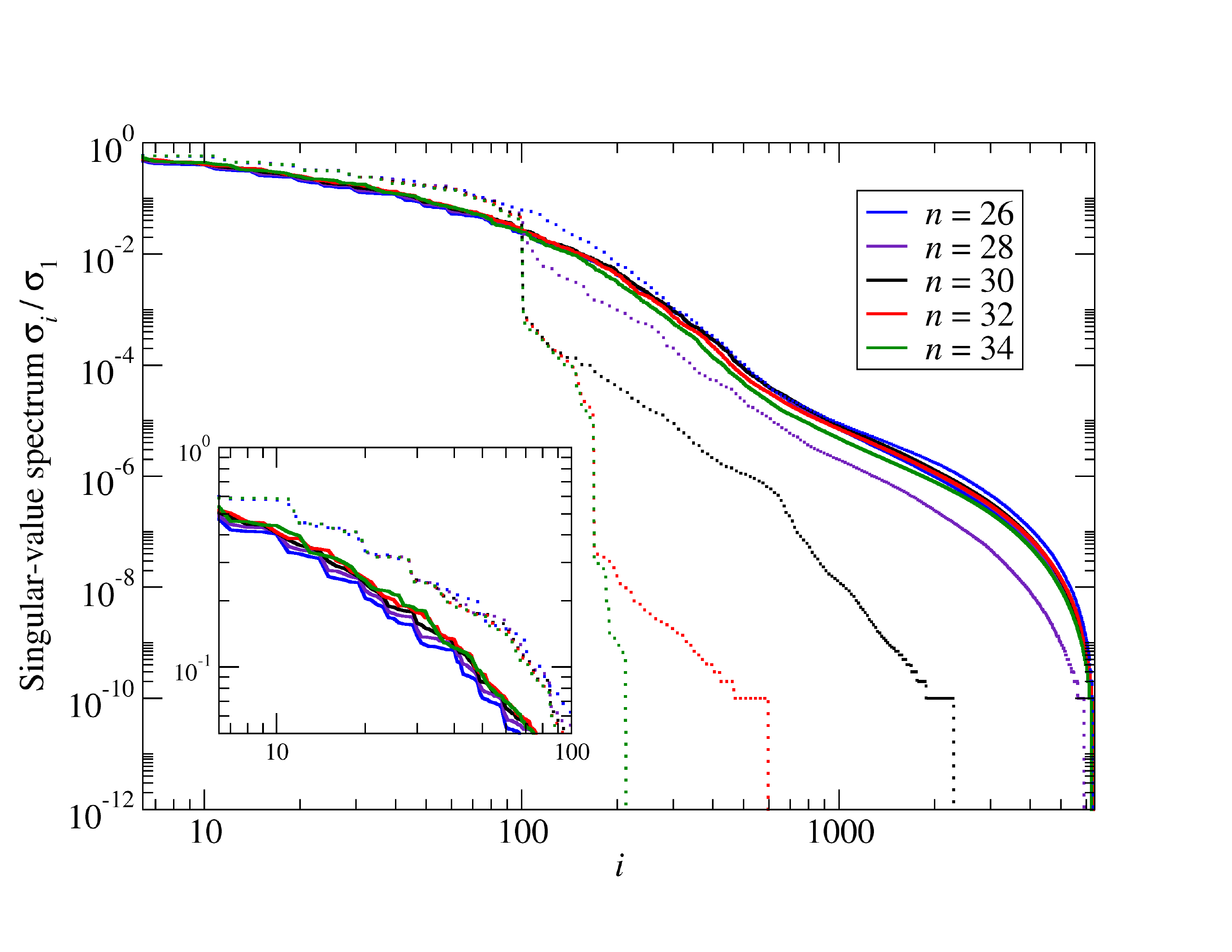}
	\caption{Singular-value spectrum obtained by the Grassmann BTRG (solid curves) with $D=80$ and $n=26, 28, 30, 32, 34$. Dotted curves represent the results provided by the Grassmann Levin-Nave TRG with $D=80$. The same color corresponds to the same renormalization step $n$. Note that as in the first line of Eq.~\eqref{eq:svd}, we can obtain $D^{2}$ numbers of singular values at each renormalization step before we introduce the low-rank approximation as in the second line of Eq.~\eqref{eq:svd}. This is the reason why the index $i$ runs beyond the bond dimension $D$. The inset graph focuses on the spectrum with $1\le i\le 100$.}
  	\label{fig:svd}
\end{figure}

\begin{figure}[htbp]
	\centering
	\includegraphics[width=0.9\hsize]{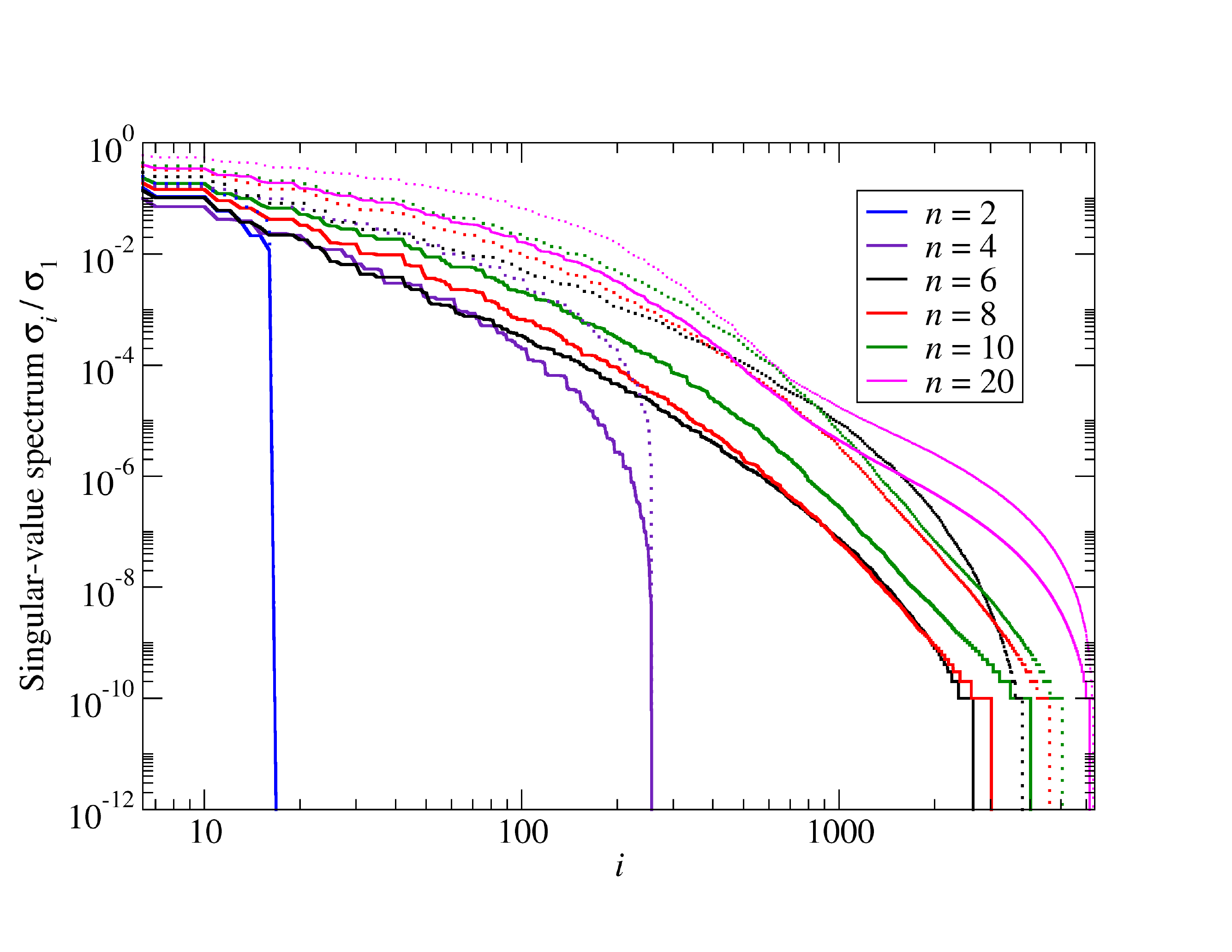}
	\caption{Singular-value spectrum obtained by the Grassmann BTRG (solid curves) with $D=80$ and $n=2, 4, 6, 8, 10, 20$. As in the figure~\ref{fig:svd}, dotted curves represent the results provided by the Grassmann Levin-Nave TRG with $D=80$. The same color corresponds to the same renormalization step $n$.}
  	\label{fig:small_svd}
\end{figure}

As another advantage demonstrated in Ref.~\cite{PhysRevB.105.L060402}, the BTRG keeps the scale-invariant structure of tensors at criticality, even though there is no explicit treatment to remove the short-range correlation in the BTRG. This feature can be confirmed by observing the stability of the singular-value spectrum under the renormalization step $n$. 
Here, we study the singular-value spectra of the Grassmann tensor with larger $n$ and smaller $n$, which are shown in Figs.~\ref{fig:svd} and \ref{fig:small_svd}, respectively. In each figure, the $y$-axis shows $\sigma_{i}/\sigma_{1}$, where $\sigma_{i}$ is the singular value in descending order and $\sigma_{1}$ is the largest one. The $x$-axis shows their subscript $i$, which runs from 1 to $D^{2}$ at most before we introduce the low-rank approximation as in the second line of Eq.~\eqref{eq:svd}. These spectra are obtained by setting $D=80$.

As the larger $n$, we choose $n=26, 28, 30, 32, 34$ in Fig.~\ref{fig:svd}. 
With $n=26$, both the Grassmann BTRG and the Grassmann Levin-Nave TRG show similar singular-value spectra. 
The former keeps the spectrum under the further coarse-graining steps, although the latter does not. 
Therefore, the Grassmann BTRG successfully produces the scale-invariant structure of the Grassmann tensors. 
Looking more closely at the larger singular values with $i\le100$, as shown in the inset graph in Fig.~\ref{fig:svd}, we can see that the singular-value spectrum obtained by the Grassmann BTRG decays more quickly than that by the Grassmann Levin-Nave TRG. Since the rapid decay of the singular value provides us with an accurate low-rank approximation, the Grassmann BTRG is superior to the Grassmann Levin-Nave TRG. 

It is also instructive to compare the singular-value spectra with the smaller $n$. We choose $n=2, 4, 6, 8, 10$ and $n=20$ in Fig.~\ref{fig:small_svd}. With $D=80$, there is no truncation up to $n=2$, so the same spectrum is obtained by both algorithms. However, the further renormalization steps result in different spectra between the Grassmann BTRG and the Grassmann Levin-Nave TRG. At every renormalization step, the singular-value spectrum obtained by the Grassmann BTRG decays much more rapidly than that by the Grassmann Levin-Nave TRG. With $n=20$, which corresponds to the lattice volume $2^{20}$ as in Figs.~\ref{fig:k_dep} and \ref{fig:d_dep}, $\sigma_{i}/\sigma_{1}$ has become less than $10^{-1}$ at $i=31$ in the Grassmann BTRG, but at $i=79$ in the Grassmann Levin-Nave TRG. In addition, $\sigma_{i}/\sigma_{1}$ has become less than $10^{-2}$ at $i=129$ in the Grassmann BTRG, but at $i=223$ in the Grassmann Levin-Nave TRG. This rapid decay of the singular-value spectrum in the Grassmann BTRG can explain the reason why the relative error of the free energy achieved by it is much smaller than the  Grassmann Levin-Nave TRG with the same bond dimension as shown in Figs.~\ref{fig:k_dep} and \ref{fig:d_dep}.

\section{Summary and outlook} 
\label{sec:summary}

We have developed the Grassmann BTRG based on the original algorithm proposed in Ref.~\cite{PhysRevB.105.L060402} and on the formalism established in Ref.~\cite{Akiyama:2020sfo}. Numerical calculations for the massless free Wilson fermion show that $k=-0.5$ is optimal as in the case of the Ising model, whose computation by the original BTRG has already been provided in Ref.~\cite{PhysRevB.105.L060402}. This means that the stationary condition in Eq.~\eqref{eq:stationary} also holds in the Grassmann tensor network. Current numerical results imply that it is the geometry of the tensor network that decides the optimal amount of bond weights, the optimal choice of the hyperparameter $k$ in other words, regardless of the particle statistics. In addition, the bond-weighting technique helps us to keep the scale-invariant structure in the Grassmann tensor network. That is, all the advantages of the BTRG claimed in Ref.~\cite{PhysRevB.105.L060402} are successfully taken over to the Grassmann BTRG. These results encourage us to apply the bond-weighting technique widely for other TRG algorithms such as the anisotropic TRG (ATRG)~\cite{Adachi:2019paf} and its Grassmann version (Grassmann ATRG)~\cite{Akiyama:2020soe}, particularly in higher dimensions.

\begin{acknowledgments}
We thank Tsuyoshi Okubo and Synge Todo for their valuable discussion.
We acknowledge the support from the Endowed Project for Quantum Software Research and Education, the University of Tokyo (\cite{qsw}). A part of the numerical calculation for the present work was carried out with ohtaka provided by the Institute for Solid State Physics, the University of Tokyo.
\end{acknowledgments}

\bibliographystyle{JHEP}
\bibliography{bib/formulation,bib/algorithm,bib/discrete,bib/grassmann,bib/continuous,bib/gauge,bib/review,bib/for_this_paper}

\end{document}